\documentclass[twocolumn]{aastex631}
\usepackage{amsmath}
\usepackage{graphicx}
\usepackage{lipsum}
\usepackage{gensymb}
\usepackage{makecell}
\usepackage{diagbox}
\usepackage{tabularx}
\usepackage{CJK}

\newcommand{\msun}{M_{\odot}}
\newcommand{\mstar}{M_\star}
\newcommand{\vstar}{v_\infty}
\newcommand{\qstar}{q_\star}
\newcommand{\istar}{i_\star}
\newcommand{\ostar}{\omega_\star}
\newcommand{\lf}{f_{i<30\degree}}
\renewcommand{\d}{\mathrm{d}}

\begin{document}
\begin{CJK*}{UTF8}{gbsn}

\title{Early Stellar Flybys are Unlikely: Improved Constraints from Sednoids and Large-$q$ TNOs}
\shorttitle{Early Sednoid-forming Stellar Flybys are Unlikely}
\shortauthors{Hu et al.}

\author[0009-0007-9015-9451]{Qingru Hu (胡清茹)}
\affiliation{Department of Astronomy, Tsinghua University, Beijing 10084, People's Republic of China}

\author[0000-0003-1215-4130]{Yukun Huang (黄宇坤)}
\affiliation{Center for Computational Astrophysics, National Astronomical Observatory of Japan, 2-21-1 Osawa, Mitaka, Tokyo 181-8588, Japan}

\author[0000-0002-0283-2260]{Brett Gladman}
\affiliation{Dept. of Physics and Astronomy, University of British Columbia, 6224 Agricultural Road, Vancouver, BC V6T 1Z1, Canada}

\author[0000-0003-4027-4711]{Wei Zhu (祝伟)}
\affiliation{Department of Astronomy, Tsinghua University, Beijing 10084, People's Republic of China}

\correspondingauthor{Yukun Huang}
\email{yhuang.astro@gmail.com}
\correspondingauthor{Wei Zhu}
\email{weizhu@tsinghua.edu.cn}

\begin{abstract}

Sedna-like objects (a.k.a. sednoids) are transneptunian objects (TNOs) characterized by large semimajor axes and exceptionally high perihelia. Their high-$q$ orbits are detached from the influence of the four giant planets and need extra perturbation to form. One hypothesis posits that close stellar flybys could have perturbed objects from the primordial scattering disk, generating the sednoid population. In this study, we run N-body simulations with different stellar encounter configurations to explore whether such a close stellar flyby can satisfy new constraints identified from sednoid (and detached extreme TNO) observation, including the low-inclination ($i < 30^\circ$) profile and primordial orbital alignment. Our results suggest that flybys with field stars are unable to generate a sufficient population, whereas flybys within the birth cluster fail to produce the primordial orbital alignment. To meet the inclination constraint of detached extreme TNOs, flybys have to be either coplanar ($\istar \sim 0^\circ$) or symmetric about the ecliptic plane ($\ostar \sim 0^\circ, \istar\sim 90^\circ$). After taking into account their occurrence rate at the early stage of the Solar System, we conclude that close-in stellar flybys ($q_\star \le 1000$~au) that satisfy all constraints are unlikely to happen ($\lesssim$5\%). Future discoveries of additional sednoids with precise orbital determinations are crucial to confirm the existence of the low-inclination tendency and the primordial alignment, and to further constrain the early dynamical evolution of the Solar System.

\end{abstract}


\section{Introduction}\label{sec:intro}

The discovery of (90377) Sedna \citep{Brown.2004} marked the first identification of a transneptunian object (TNO) with large semimajor axis ($200<a<1500$ au) and exceptionally high perihelion distance ($q>60$ au)\footnote{The cuts of semimajor axis and perihelion distance for sednoids vary greatly in the literature: for example, \citet{Wajer.2024} used $100<a<1600$ au and $q>60$ au, while \citet{Huang.2023t} used $200<a<1500$ au and $q>60$ au; here we adopt the latter.}. This discovery was soon followed by the identification of two more Sedna-like objects: 2012 VP$_{113}$ \citep{Trujillo.2014} and (541132) Leleakuhonua (2015 TG$_{387}$) \citep{Sheppard.2019}. These objects, also collectively known as sednoids, are distinguished by their detached perihelia and stable orbital evolutions for the age of the Solar System \citep[e.g.,][]{2021ApJBatygin,2024MNRASHadden,2025arXivBelyakov}.

Sednoids are thought to have formed through the synergy of early planetary scatterings of planetesimals in the primordial disk, which increased their semimajor axes beyond a few hundred au, and additional perturbations (beyond the four known giant planets) elevating their perihelia and detaching them from Neptune's scattering zone. The orbital detachment (or $q$-lifting) mechanism remains a hotly debated topic. Hypotheses include a short-lived rogue planet made in the early Solar System \citep{Gladman.2006, Huang.2022ml}, a still present solar companion/planet \citep{2002Gladman, Gomes.2006, PNsim_Lykawka2008, Batygin.2016}, close stellar flybys in the Sun's birth cluster \citep{Morbidelli.2004, Kenyon.2004, Brasser2006, Nesvorny.2023, Pfalzner.2024}, an early solar binary \citep{Raush.2024}, and stellar flybys during solar migration in the Milky Way \citep{Kaib.2011}. There is an alternative scenario where sednoids could have been captured from the outer disk of a passing star \citep{Morbidelli.2004,Kenyon.2004,2015MNRAS.453.3157J}.

Recent work by \citet{Huang.2024s} revealed that the longitudes of perihelion ($\varpi = \Omega + \omega$) of the three known sednoids were tightly clustered $\approx$4.5~Gyr ago at a longitude of $\varpi_{-4.5\text{Gyr}} \approx 200^\circ$ with a circular standard deviation of only $8\degree$. This ``primordial orbital alignment'', which is unrelated to the observational biases affecting the current distant Kuiper Belt \citep{Shankman.2017, Napier.2022}, suggests that an initial dynamical event may have imposed this specific apsidal orientation on the early sednoid population. Subsequent orbital evolution would then be mainly driven by the linear secular precession induced by the four giant planets and, to a lesser extent, by galactic tides.

If future discoveries of additional sednoids confirm this primordial alignment, it would impose valuable constraints on not only the current state of the outer Solar System but also mechanisms responsible for the formation of sednoids. In particular, models involving the current presence of a distant planet is likely incompatible with the alignment. \citet{Huang.2024s} inspected both the rogue planet and the close stellar flyby scenarios as possible explanations for the primordial alignment. The rogue planet model, where an additional massive planet elevates sednoid perihelia before being ejected, naturally produces the apsidal alignment along its own apsidal line. The preliminary study of close stellar flybys ($q_\star = 300$~au and $\vstar = 1$~km/s) demonstrated they could also elevate sednoid orbits, although such an encounter did not easily generate strongly clustered longitudes. The rogue planet and the stellar flyby scenarios also differ in the produced inclination distribution; while the former produces a low-$i$ profile \citep{Gladman.2006, Huang.2023t}, the latter tends to produce a heated inclination distribution, including highly inclined orbits ($i>60^\circ$) and even retrograde orbits ($i>90^\circ$) \citep{Brasser.2014,Wajer.2024}.

In this study, we extend the previous study \citep{Huang.2024s} by conducting comprehensive numerical simulations to assess whether close stellar flybys can satisfy all the observational constraints identified from the sednoid (and detached extreme TNO) population, including the low-$i$ profile and the primordial orbital alignment of sednoids.

\section{Simulation Setup}\label{sec:setup}

The simulation setup for a single stellar flyby is as follows. Planetesimals from the primordial scattering disk (implemented as massless test particles) are numerically integrated with the Sun and a passing star, using the \texttt{IAS15} integrator \citep{reboundias15} in the \texttt{REBOUND} N-body code \citep{rebound}. The four giant planets are not included, because the flyby timescale ($\lesssim10^3$ yr) is generally shorter than the orbital periods of $a \gtrsim 200$~au planetesimals ($\gtrsim3\times10^3$ yr). Consequently, the longer-term scattering effect of the giant planets, which acts on the planetesimals over multiple orbits, can be safely ignored during the stellar flyby. Once generated by the stellar flyby, the semimajor axes, perihelia, and inclinations of sednoids remain largely stable over 4.5 Gyr under the gravitational effects of known Solar System objects \citep[e.g.,][]{Sheppard.2019, Huang.2024s}. To further validate our approach of excluding the four giants, we performed representative stellar flyby simulations (as in Figure~\ref{fig:a-q-i}) that include the four giants, and found that the absolute variations in the sednoid implantation efficiency ($\eta$) and the fraction of $i<30\degree$ sednoids ($\lf$; defined in Section~\ref{subsect:overall}) are $<$2\%.

The initial conditions of the primodial scattering disk are generated as follows: The perihelion distances ($q$) are sampled from a uniform distribution of $\mathcal{U}[5, 25]$ au, corresponding to the region where the four giant planets actively scatter planetesimals during their early migration phase \citep{Nesvorny.2016}. While the actual distribution of perihelia in the primordial scattering disk likely has more structure, somewhat concentrated at the giant planet orbital distances, our simplified uniform distribution is sufficient for this study, since the final sednoids we analyze have $q > 60$ au. The semimajor axes ($a$) are sampled from $\d N / \d a \propto a^{-1.5}$ with $a\sim(100, 1500)$ au, which is  the steady-state distribution of a planet-scattering disk \citep{1980Yabushita, Levison.1997, Huang.2023t}. The inclinations ($i$) are sampled from $\d N / \d i \propto \mathcal{N}[0, 15](^{\circ}) \times \sin i $ obtained from OSSOS survey for detached TNOs \citep{Beaudoin.2023}. The arguments of perihelion $\omega$, the longitudes of ascending node $\Omega$, and the mean anomalies $\mathcal{M}$ are all randomly sampled from 0 to $2\pi$.

Previous studies employing realistic stellar cluster models \citep[e.g.,][]{Brasser2006,Wajer.2024} have shown that the implantation of sednoids is dominated by the strongest stellar encounter. To better understand the effects of a single close stellar flyby, we therefore focus on discrete, representative cases with: \textbf{(1)} two different perturber masses ($M_\star = \msun$, a sun-like star, and $M_\star = 0.3 \ \msun$, the typical stellar mass in a young embedded cluster, \citealt{kroupa2002IMF}); \textbf{(2)} four different relative velocities at infinity $\vstar = \{1, 3, 10, 30\}$ km/s, where the former two are the typical velocity dispersions of young clusters \citep{Brasser2006}, and the latter two are those of field stars; \textbf{(3)} four closest approach distances $q_\star = \{ 300, 500, 700, 1000\}$~au, where the stability of giant planets \citep{adams2010birth} and the cold classical Kuiper Belt objects \citep{Batygin.2020jr} imposes a lower bound of $\qstar \gtrsim 240$ au while the implantation of sednoids in a stellar flyby scenario imposes an upper limit of $\qstar \lesssim 800$ au \citep{Morbidelli.2004}. 

Besides the shape of the hyperbola defined by $M_\star$, $\vstar$ and $q_\star$, the orientation of the stellar flyby relative to the Solar System's plane also matters. Given the axisymmetry of the scattering disk about the $z$-axis, the longitude of ascending node of the hyperbolic trajectory is fixed to 0. Given the two-fold symmetry of the argument of periapsis, we only consider four different $\omega_\star = \{0^{\circ}, 45^{\circ}, 90^{\circ}, 135^{\circ}\}$. Lastly, six different inclinations $i_\star = \{0^{\circ}, 36^{\circ}, 72^{\circ}, 108^{\circ}, 144^{\circ}, 180^{\circ}\}$ are considered to account for both prograde and retrograde flybys.  This yields 768 unique stellar encounter configurations, and each simulation set contains 10,000 test particles in the primordial scattering disk. Each simulation begins and ends with the passing star at a distance of 30,000 au from the Sun.

\section{Analysis and Results}\label{sect:results}

\subsection{Overall Distributions in $a$-$q$-$i$ Space}\label{subsect:overall}

\begin{figure*}[t]
    \centering
    \includegraphics[width=\linewidth]{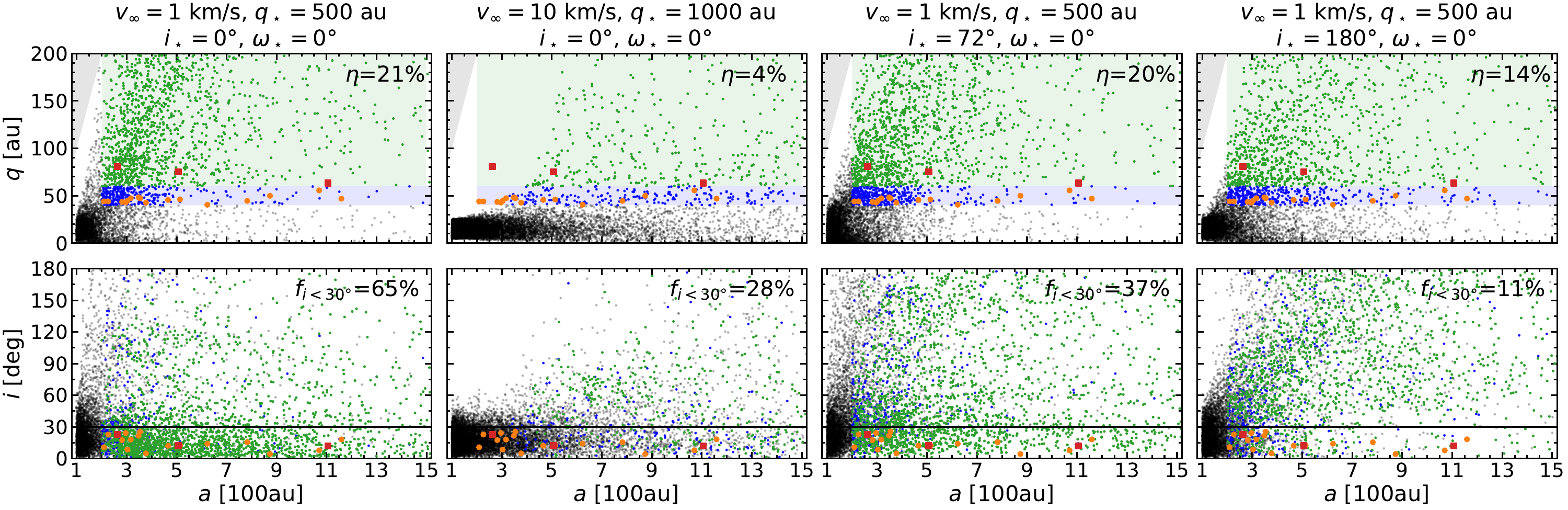}
    \caption{The $a$-$q$-$i$ distribution of all TNOs produced by four example stellar encounters for a flyby $M_\star=1M_\odot$. In the upper panels, the definitions of sednoids ($200<a<1500$~au, $q>60$~au) and detached extreme TNOs (eTNOs, $200<a<1500$~au, $q>40$~au)  are emphasized by the green and blue shaded regions. Synthetic sednoids, eTNOs and other TNOs are marked as green, blue and gray small dots. The red square dots are three observed sednoids, and the orange circle ones are observed eTNOs excluding sednoids. In the lower panels, the black lines mark out $i=30^{\circ}$. The implantation efficiency of sednoids ($\eta$) and the fraction of $i<30\degree$ sednoids ($\lf$) are annotated in the upper right corners for clarification.}
    \label{fig:a-q-i}
\end{figure*}

We begin our analysis by examining the orbital distributions in semimajor axis ($a$), perihelion distance ($q$), and inclination ($i$) of TNOs generated from four representative sets of stellar encounter parameters (see Figure~\ref{fig:a-q-i}). This figure illustrates the influence of three key parameters: the relative stellar velocity at infinity ($\vstar$), the encounter perihelion distance ($q_\star$), and the encounter inclination ($i_\star$). 
Because sednoids—taken here as TNOs with $200 < a < 1500$~au and $q > 60$~au—have exceptionally high perihelia, their post-formation evolution is largely independent of perturbations from the giant planets (except for secular precessions). This assumption enables a direct comparison between our synthetic sednoid population and the observed one in the $a$-$q$-$i$ space\footnote{Objects with $1,000 < a < 1,500$~au experience small oscillations in $q$ and $i$ due to galactic tides \citep{Sheppard.2019}, which we ignore here.}.

Our simulations confirm that the $a$-$q$ distribution of synthetic sednoids is highly sensitive to $\vstar$ and $q_\star$. Firstly, we introduce the implantation efficiency of sednoids ($\eta$), which is defined by dividing the number of synthetic sednoids by the initial number (10,000) of test particles in the primordial scattering disk, as a key feature of the $a$-$q$ distribution. As illustrated in the upper panels of Figure~\ref{fig:a-q-i}, increasing $\vstar$ from 1~km~s$^{-1}$ to 10~km~s$^{-1}$ and expanding $q_\star$ from 500~au to 1000~au causes the implantation efficiency to drop significantly—from roughly 21\% down to 4\%. In contrast, variations in the encounter inclination $i_\star$ (ranging from prograde $i_\star = 0^\circ$, to retrograde $i_\star = 180^\circ$) have a far lesser effect, with implantation efficiency consistently remaining above 10\%. A more detailed discussion of the sednoid implantation efficiency is provided in Section~\ref{subsect:implant}.

\begin{figure*}
    \centering
    \includegraphics[width=\linewidth]{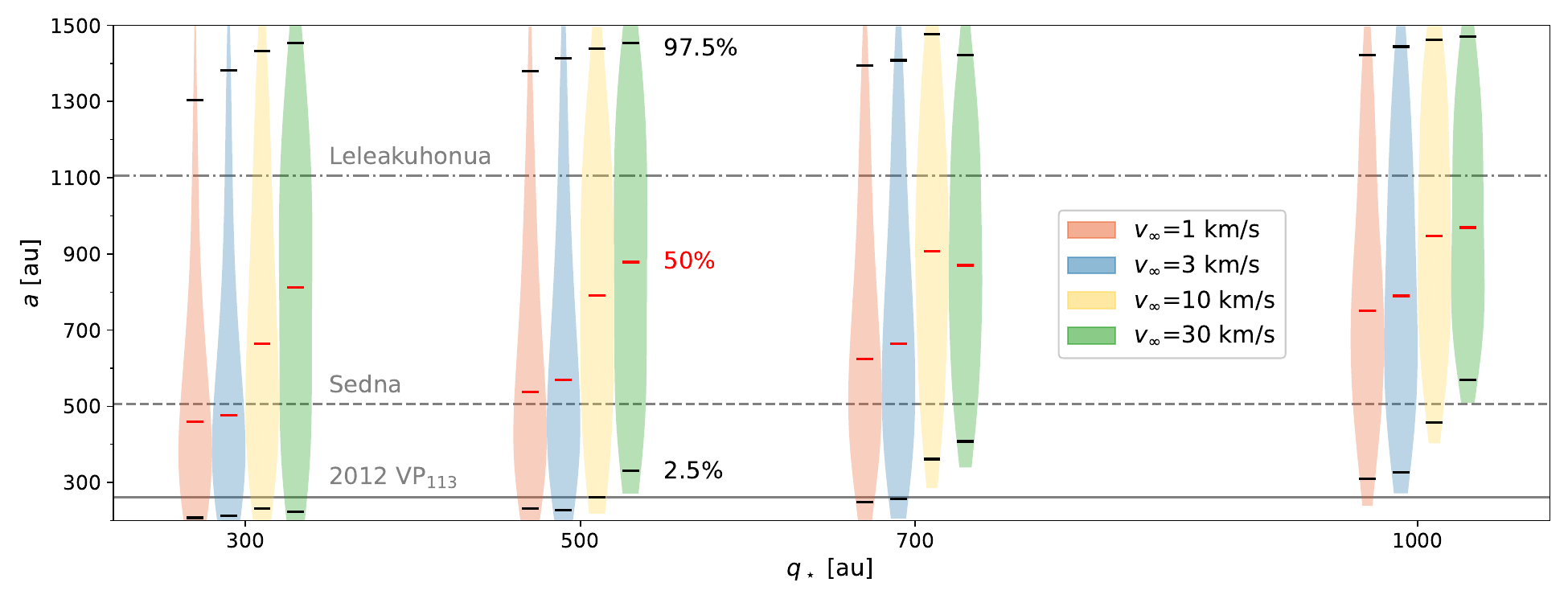}
    \caption{Distributions of sednoid semimajor axis from simulations for different $\vstar$ and $q_\star$, with both $i_\star$ and $\omega_\star$ fixed to 0. The red short line indicates the median value of the distribution, and the black short lines show the 2.5\% and 97.5\% percentiles. The semimajor axes of three observed sednoids are marked out as gray lines.}
    \label{fig:a-range}
\end{figure*}

A successful stellar flyby simulation should at least encompass the $a$-$q$-$i$ range of the observed large-$a$ TNOs. While the coverage in $q$ and $i$ space is always sufficient (see Figure~\ref{fig:a-q-i}), the high-speed, high-periastron simulations (specifically, those with $\vstar \geq 10$~km~s$^{-1}$ and $q_\star \geq 1000$~au) do not produce the innermost sednoid, 2012 VP$_{113}$ ($a_{\text{VP}_{113}}\approx 262$ au). As Figure~\ref{fig:a-range} demonstrates, the median and 2.5\% percentile ($a_{\text{sim}}^{2.5\%}$) of the semimajor axis distribution of synthetic sednoids are in general positively correlated to $\vstar$ and $\qstar$, while the 97.5\% percentile ($a_{\text{sim}}^{97.5\%}$) remains relatively unaffected. Under extreme cases, every encounter scenario with $q_\star = 1000$~au is unable to match the orbit of 2012 VP$_{113}$ within the 95\% confidence interval.

\subsection{Implantation Efficiency}\label{subsect:implant}

\begin{figure*}[t]
    \centering
    \includegraphics[width=\linewidth]{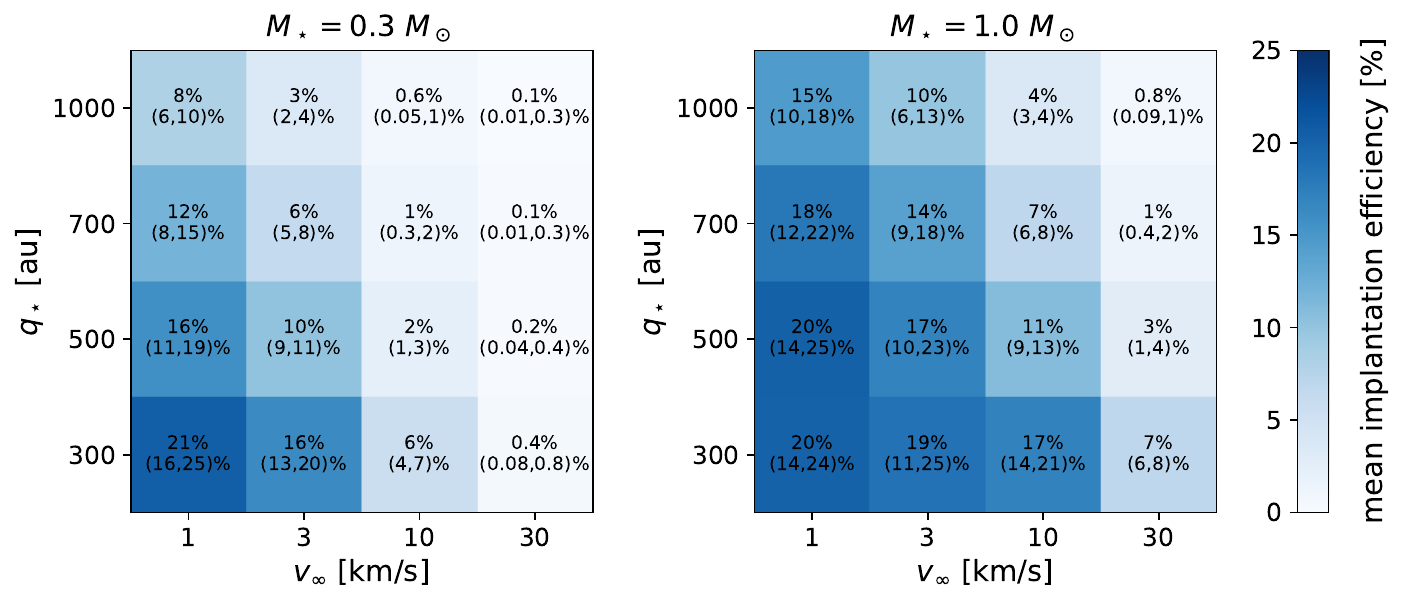}
    \caption{The implantation efficiency of sednoids for $M_\star = 0.3 \msun$ and $M_\star = 1 \msun$, projected onto the $(\vstar, q_\star)$ space. The mean implantation efficiency across 24 possible sets of ($i_\star, \omega_\star$) for fixed $(\vstar, q_\star)$ is marked above the parentheses and also represented by the grid color，while the minimum and maximum values are marked in the parentheses.}
    \label{fig:implantation-rate}
\end{figure*}

Before diving into the analysis of orbital properties of synthetic sednoids, we investigate the implantation efficiency of sednoids ($\eta$) under different stellar flybys and estimate the minimum $\eta$ required by current observation.

Figure~\ref{fig:implantation-rate} presents the statistical distribution of implantation efficiency, showing extrema and mean values for fixed stellar velocity and perihelion distance ($\vstar$, $q_\star$) across all angular parameters ($\omega_\star$, $i_\star$). Systematic analysis reveals mean implantation efficiency of $\bar{\eta} = 6.4\%$ for $M_\star=0.3M_\odot$ versus $\bar{\eta}=11.4\%$ for $M_\star=1M_\odot$, with a clear positive correlation between stellar mass and implantation efficiency for every $(\vstar,\qstar)$ pair. Examining 1$M_\odot$ encounters, the efficiency exhibits a factor of 30 decline from $\eta=20.2\%$ to $\eta=0.8\%$ as $\vstar$ increases from 1 km/s to 30 km/s and $q_\star$ from 300 au to 1000 au. This inverse relationship confirms that optimal perturbation occurs during low-velocity, close-proximity encounters—conditions favoring strong gravitational perturbation and enhanced dynamical exicitation.

Recent sednoid-forming simulations using an embedded cluster scenario \citep[e.g.,][]{Wajer.2024} and a rogue planet model \citep[e.g.,][]{Huang.2023t} found that roughly $2\times10^5$ planetesimals with diameters $D>100\,$km ($H_r>8.3$) would be implanted into the sednoid region. Based on the latest size distributions for the hot TNO population \citep[e.g.,][]{Petit.2023,Ormel.2025}, this number translates to roughly $\sim$$20$ Sedna-sized objects, $\sim$$100$ VP$_{113}$-sized objects, and $\sim$$1000$ Leleakuhonua-sized objects. Given the implantation efficiency of $\approx$1\% from a primordial outer disk of $\sim$$20\,M_\oplus$ \citep{Nesvorny.2018}, the inferred total sednoid mass is approximately $\sim$$0.2\,M_\oplus$. In contrast, our simulations do not start from a low-$e$, low-$i$ planetesimal disk; rather, we initialize a steady-state \textit{scattering} disk characterized by $\mathrm{d}N/\mathrm{d}a \propto a^{-1.5}$. Under the optimistic assumption that all planetesimals from the outer disk are immediately scattered out, the fraction of particles in the $a\sim(100, 1500)$~au range relative to those in $a\sim(5, \infty)$~au is $\approx$$20\%$. This value represents an upper bound, since many planetesimals are not instantaneously scattered by Jupiter---indeed, a significant fraction continues to experience scattering by Neptune over a much longer ($\sim$100 Myr) timescale \citep{Nesvorny.2016}.

Therefore, we deem stellar encounters with $\eta<5\%$ to be inconsistent with current observations, as they would necessitate an unrealistically massive primordial disk to generate the observed sednoid population. Figure~\ref{fig:implantation-rate} illustrates that this constraint effectively eliminates high-velocity and distant stellar encounters from consideration. Specifically, for perturber masses of $0.3 M_\odot$ and $1 M_\odot$, encounters with velocities greater than and equal to 10 km/s and 30 km/s respectively, are generally inefficient at producing sednoids.

\subsection{Fraction of $i<30^{\circ}$ Sednoids}\label{subsect:low-i}

\begin{figure*}[t]
    \centering
    \includegraphics[width=\linewidth]{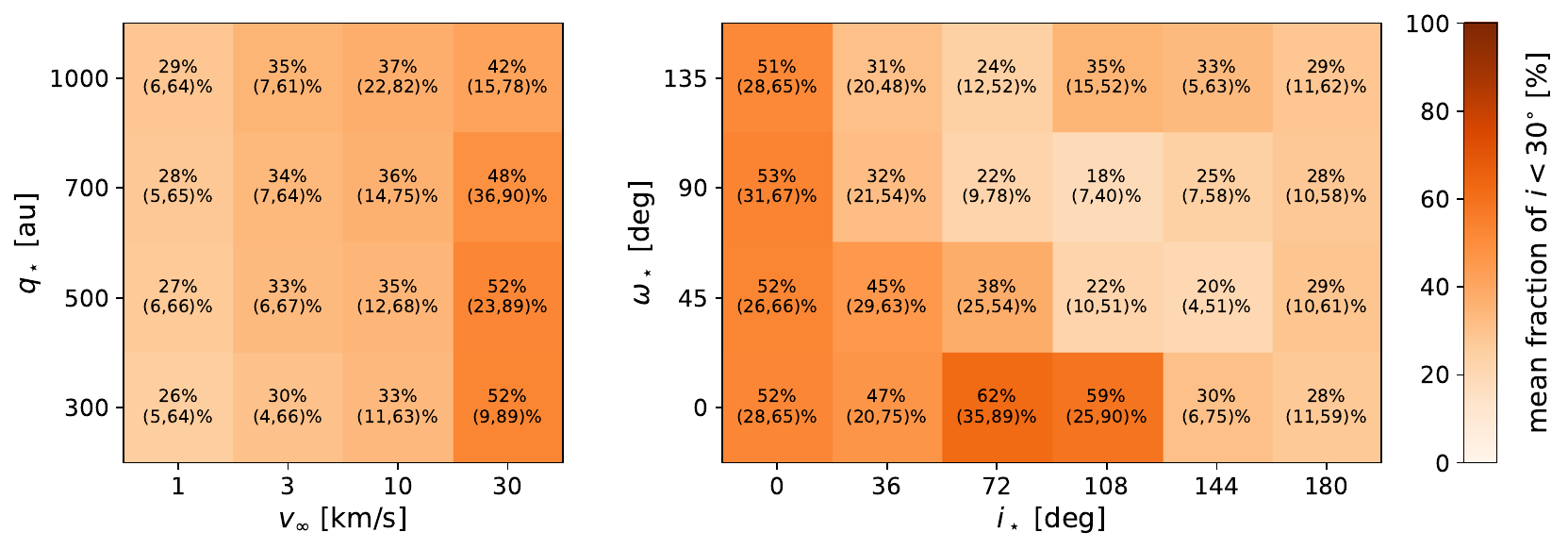}
    \caption{The fraction of $i<30\degree$ sednoids for stellar encounters with $M_\star = 1 \msun$, projected onto the $(\vstar, q_\star)$ (left panel) and $(i_\star, \omega_\star)$ (right panel) parameter spaces.
    The mean $i<30\degree$ fraction across the remaining stellar parameters is marked above the parentheses and also represented by the grid color, while the minimum and maximum values are marked in the parentheses, same as Figure~\ref{fig:implantation-rate}.}
    \label{fig:low-inc-frac}
\end{figure*}

The inclination distribution of distant TNOs provides a crucial constraint on their formation mechanism. Notably, all three known sednoids have $i<30^\circ$. This low-inclination trend extends to the broader population of extreme trans-Neptunian objects (eTNOs, defined as objects with $200<a<1500$~au and $q>40$~au \footnote{The cuts of semimajor axis and perihelion distance for detached eTNOs also vary in the literature.}; \citealt{PN_Batygin2019}). Among the 19 known eTNOs, including the three sednoids, all have $i<30^\circ$ (see Figure~\ref{fig:a-q-i} and Table~\ref{tab:survey}). 
This prevalence of low inclinations among the most distant TNOs was initially noted by \citet{Gladman.2006} and has been recently reaffirmed by \citet{Wajer.2024}, who found that formation scenarios involving embedded clusters struggle to reproduce the observed consistently low sednoid inclinations. 
Given this potential inconsistency between observations and stellar flyby simulations, we revisit this aspect to assess the viability of the passing star hypothesis. Only in this section, where the inclination distribution is studied, do we utilize the entire eTNO sample. This is because, although Neptune-driven diffusion causes random walks in semimajor axis for $q < 60$~au eTNOs, their orbital inclinations remain largely unchanged \citep{2024MNRASHadden}.

As shown in Figure~\ref{fig:a-q-i} and \ref{fig:low-inc-frac}, the low-$i$ fraction of produced sednoids is highly correlated with the orientation of the stellar encounters. In particular, encounters with $i_\star=0^\circ$ produce the highest fraction of $i<30^\circ$ sednoids ($\lf \gtrsim$ 65\%). As $i_\star$ increases, the low-$i$ fraction generally decreases. For instance, under fixed $\omega_\star=90^\circ$, the mean $\lf$ decreases from 53\% for $i_\star=0^\circ$ to 18\% for $i_\star=108^\circ$. This trend, however, is not true for vertical flybys symmetrical to the orbital plane (i.e., with $i_\star\approx90^\circ$ and $\omega_\star=0^\circ$). Such flybys, despite of being vertical, maintain $\lf \gtrsim 30\%$, as the vertical impulse of the passing star symmetrically cancels about the ecliptic plane (see Figure~\ref{fig:low-inc-frac}'s fourth row in the right panel and Figure~\ref{fig:a-q-i}'s third column for an example). The left panel of Figure~\ref{fig:low-inc-frac} illustrates the dependence of $\lf$ on $\vstar$ and $q_\star$. Flybys with lower $\vstar$ and smaller $q_\star$ typically result in a more heated $i$ distribution, leading to a lower $\lf$.

It is essential to stress that although the eTNOs were discovered by various surveys with different telescopes and selection functions (see detailed discussions in Appendix \ref{appen:1}), we can still assess the consistency with stellar flyby models using a forward-biasing simulation. The simulation is similar to the OSSOS survey simulator \citep{Lawler.2018}, where one projects an intrinsic model (in our case, particles generated by stellar flybys) onto the sky, with random mean anomalies and a reasonable $H_r$ distribution \citep{Petit.2023}. By assuming all past surveys (e.g., \citealt{Trujillo.2014, Bannister.2017, Sheppard.2019, Bernardinelli.2022}) that discovered eTNOs have an average off-ecliptic coverage of $\pm30^\circ$ around the ecliptic plane (justified in Appendix \ref{appen:1}), a forward-biased synthetic sample is generated. The synthetic $i$ distribution strongly depends on the intrinsic $i$ distribution, and is less sensitive to the assumed $a$, $q$, and $H_r$ distributions \citep{Beaudoin.2023}. In contrast, similar consistency tests for $a$ and $q$ are hampered by the need to accurately compile the magnitude limits and coverages of independent surveys.

\begin{figure}
    \centering
    \includegraphics[width=0.95\linewidth]{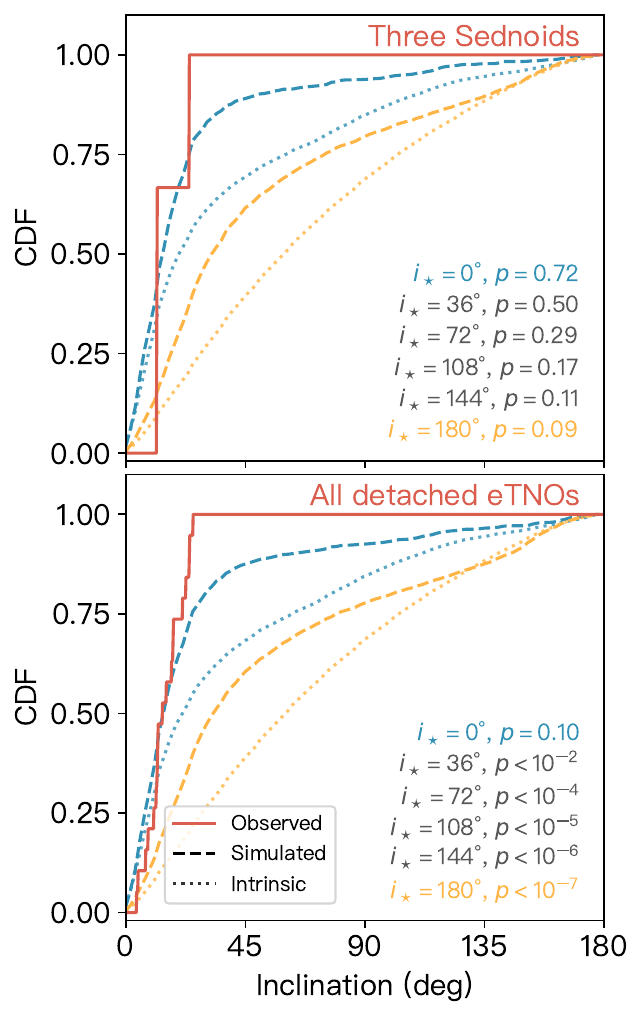}
    \caption{Inclination distributions of synthetic sednoids ($q>60$~au, upper panel) and detached eTNOs ($q>40$~au, lower panel) produced by stellar flybys with $i_\star=0^\circ$ (blue) and $i_\star=180^\circ$ (orange) for $\vstar=1$~km/s, $\qstar = 300$~au, and $M_\star=1M_\odot$. The intrinsic distributions (which should not be compared to the real detections) are shown with dotted lines, and the real observed ones are shown with solid red lines. The synthetic simulated distributions (dashed lines) are the biased distributions produced by the simulated surveys with a typical coverage of $\pm30^\circ$ around the ecliptic plane. The $p$-values of the KS test between the simulated detections and the observed samples are given. Distributions for the four othr $\istar$ cases are not plotted (for clarity), but their intrinsic and synthetic curves fall between blue and orange ones.}
    \label{fig:i_dist_v3}
\end{figure}

Figure~\ref{fig:i_dist_v3} shows the cumulative $i$ distributions for stellar encounters with $\vstar=1$~km/s, $\qstar=300$~au, and $M_\star=1M_\odot$. The inclination of the passing star significantly affects the $i$ distribution, with the strongest excitation occurring at $i_\star=180^\circ$ and weakest at $i_\star=0^\circ$ (with the notable exception of $\omega_\star\approx0^\circ$ flybys as explained previously). When compared to synthetic detections (dashed lines) using the Kolmogorov--Smirnov (KS) test, the three sednoids alone cannot provide meaningful constraints. However, when compared to the entire eTNO sample, the KS test reveals that off-plane flybys ($i_\star\geq36^\circ$) generally produce overly excited $i$ distributions. 
This trend persists across different flyby parameters, suggesting that to remain consistent ($p$-value $>0.05$) with the unheated $i$ distribution of eTNOs, the intrinsic $\lf$ exceeds 50\%. Given the similar $i$ distributions of eTNOs and sednoids (see Figure~\ref{fig:a-q-i}), which likely formed through the same process, we use this 50\% threshold to constrain our flyby simulations. 

Overall, these results suggest that only stellar flybys with favorable orientations (notably $i_\star\approx0^\circ$, or $\omega_\star=0^\circ$ with $i_\star\sim90^\circ$) can produce a significant fraction of low-inclination TNOs consistent with observations. This strong correlation between the flyby orientation and the resulting inclination distribution offers a promising avenue to constrain the dynamical history of the outer Solar System. 


\subsection{Primordial Orbital Alignment}\label{subsect:alignment}

\begin{figure*}
    \centering
    \includegraphics[width=\linewidth]{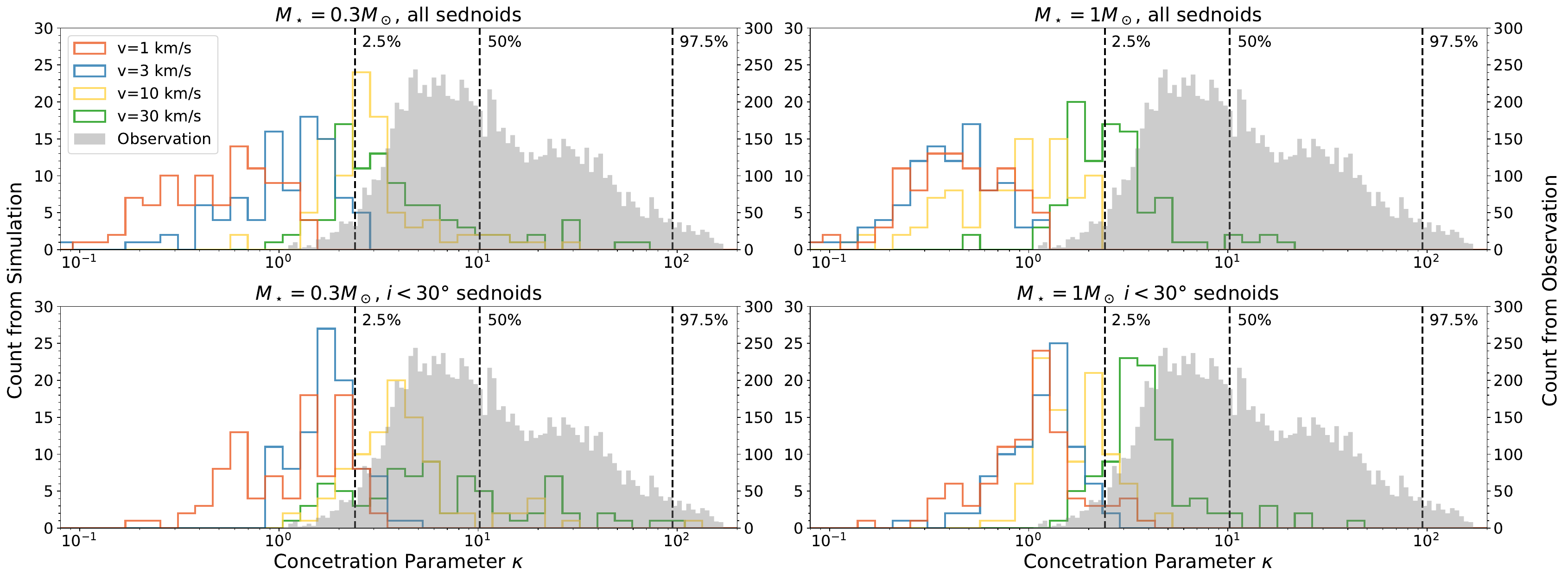}
    \caption{The histograms of the concentration parameter $\kappa$ derived from both simulation (colored steps) and observation (gray-filled steps) for stellar masses of $M_\star=0.3 M_\odot$ (left panels) and $M_\star=1 M_\odot$ (right panels). The upper panels present results using all simulated sednoids for the calculation of $\kappa$, whereas the lower panels restrict the analysis to synthetic sednoids with inclinations $i < 30\degree$. The 95\% confidence interval (ranging from 2.5\% to 97.5\%) of the observed $\kappa$ distribution is indicated by black dashed lines.}
    \label{fig:kappa_hist}
\end{figure*}

Recently, \citet{Huang.2024s} discovered an intriguing phenomenon: when integrated backward for $\sim$4.5~Gyr, the orbits of the three known sednoids all pointed in the same direction, a feature dubbed the ``primordial orbital alignment''. If confirmed by future sednoid discoveries, this alignment would provide a valuable constraint on sednoid formation mechanisms. Assuming the primordial orbital alignment is real, we examine whether the stellar flyby hypothesis is consistent with this emerging constraint.

We employ the concentration parameter $\kappa$ from the 3D von Mises--Fisher (vMF) distribution \citep{Fisher.1993, Matheson.2023} of perihelion directions on the unit sphere as a quantitative measure of orbital alignment. This approach, also utilized by \citet{Pichierri.2025} in their analysis of current orbital clustering, allows for a robust statistical comparison between simulated and observed sednoid populations.

The 3D vMF distribution is defined as
\begin{equation}
    f_{\text{vMF, 3D}}(\boldsymbol{x};\boldsymbol{\mu},\kappa)= \frac{\kappa}{4\pi \sinh \kappa}\exp(\kappa\boldsymbol{\mu}\cdot \boldsymbol{x}),
\end{equation}
where $\boldsymbol{x}$ is a random unit vector, $\boldsymbol{\mu}$ is the mean direction and $\kappa$ is the concentration parameter. This distribution is the extension of the von Mises circular distribution, the analogy of the normal distribution on a circle, to a sphere in 3D space.

For each set of synthetic sednoids, we fit a vMF distribution to their primordial perihelion directions (i.e., the orientation of their Laplace--Runge--Lenz vectors after the stellar flyby) using \texttt{scipy} and extract the concentration parameter $\kappa$. A higher $\kappa$ indicates a stronger clustering of perihelion directions, corresponding to a more pronounced primordial alignment. To account for the uncertainty in the primordial $\kappa$ contributed by the \textit{current} orbital uncertainties in the three observed sednoids (especially Leleakuhonua, whose $\delta a \approx 200$~au), we compute the distribution of $\kappa$ in Figure~\ref{fig:kappa_hist} by randomly selecting one orbit from the 1,000 clone orbits for each of the three real sednoids, repeating this process 10,000 times. These clone orbits are computed from our updated simulations similar to those in \citet{Huang.2024s}, where the objects are integrated backward 4.5 Gyr accounting for the four giant planets and galactic tides.
The resulting $\kappa$ distribution is represented by the gray-filled histogram in Figure~\ref{fig:kappa_hist}, with the 95\% confidence interval highlighted by black dashed lines. We adopt this confidence interval, ranging from $\kappa^{2.5\%}_{\text{obs}} = 2.4$ to $\kappa^{97.5\%}_{\text{obs}} = 94.8$, with a median value $\kappa^{50\%}_{\text{obs}} = 10.2$, as the valid range to potentially reject certain stellar flyby simulations.

As shown in Figure~\ref{fig:kappa_hist}, most low-velocity stellar encounters ($v_\star = 1, 3$ km/s) fail to produce a strong alignment ($\kappa > \kappa^{2.5\%}_{\text{obs}}$), regardless of whether one considers all synthetic sednoids or only those with $i < 30^\circ$. In contrast, high-velocity stellar encounters ($v_\star = 10, 30$ km/s) produce more clustered perihelia, which exhibit significant overlap between the produced $\kappa$ and the observed 95\% confidence interval, particularly for $M_\star = 0.3 M_\odot$ and $i < 30^\circ$ sednoids.

To complement this analysis, we perform a statistical comparison using the Monte Carlo method to account for bias arising from small-number statistics. For each set of stellar encounter parameters, we compare $\kappa$ derived from: (1) three synthetic sednoids randomly selected from simulations, and (2) three real sednoids with each orbit randomly chosen from the 1,000 clone orbits. Repeating this procedure 1,000 times, we find that for $v_\star =$ 1 or 3 km/s, there is a $\approx$17\% chance that the simulated $\kappa$ exceeds the observed values, while for $v_\star =$ 10 or 30 km/s, this probability increases to $\approx$33\%.

In conclusion, if the primordial alignment persists with more sednoid discoveries, it would disfavor low-velocity stellar flybys ($v_\star \lesssim 3$ km/s), the kinds of speed thought to occur in the postulated birth cluster of the Sun. Combining these findings with other constraints discussed in previous sections, one can impose stringent limitations on the stellar flyby hypothesis.

\subsection{Summary of Results}\label{subsect:summary}

Based on the three currently known sednoids (Sedna, 2012 VP$_{113}$, and Leleakuhonua), we establish the following four observational constraints:
\begin{itemize}
    \item Implantation Efficiency ($\eta>5\%$): The minimum implantation efficiency of sednoids from the primordial scattering disk must exceed 5\% (Section~\ref{subsect:implant});

    \item Semimajor Axis Range ($a_{\text{sim}}^{2.5\%} < a_{\text{sed}} < a_{\text{sim}}^{97.5\%}$): The orbits of known sednoids should be encompassed within the 95\% confidence interval of the simulated sednoid $a$ distribution (Section~\ref{subsect:overall});

    \item Low-inclination Fraction ($f_{i<30^\circ} > 50\%$): Over 50\% of the generated sednoid population must maintain inclinations below $i<30^\circ$, consistent with observational statistics derived from detached eTNOs, which likely formed in the same dynamical process as sednoids (Section~\ref{subsect:low-i});

    \item Primordial Clustering Range ($\kappa^{2.5\%}_{\text{obs}} < \kappa_{\text{sim}} < \kappa^{97.5\%}_{\text{obs}}$): The $\kappa_\text{sim}$ parameter, quantifying orbital clustering for simulated $i<30^\circ$ sednoid populations, must fall within the 95\% confidence interval of the observed value (Section~\ref{subsect:alignment}).
\end{itemize}

Applying these four constraints, we find that the viable parameter space is reduced to only 29 of the 768 stellar encounter configurations, among which 27 are for extremely close ($\qstar = 300$ au) stellar encounters with field stars ($\vstar=10$ km/s with $\mstar=0.3 \msun$, or $\vstar=30$ km/s with $\mstar=1 \msun$), while the remaining 2 solutions imply close ($\qstar = 300$ au) stellar encounters within the Sun's birth cluster ($\vstar=3$ km/s with $\mstar=0.3 \msun$).

For the first 27 solutions, one is concerned that such close encounters with field stars are rare in the early days of the Solar System in real scenarios. If we take a velocity dispersion of $\vstar=10$ km/s and assume that the number density of field stars is $n_\star\approx 0.1$ pc$^{-3}$ \citep{adams2010birth}, then the average number of close flybys with $\qstar=300$ au over $t=500$ Myr is $N = n_\star \vstar \qstar^2 t \approx 0.003$, which indicates that the probability of this kind of flyby is $<1\%$.
For the later ones, although the occurrence rate is higher for such close encounters in embedded clusters, these 2 solutions all have a fine-tuned configuration of $\ostar=0\degree$ with $\istar\sim 90\degree$, due to the low-inclination fraction constraint. The probability for $\ostar < 22.5\degree$ and $72\degree < \istar < 108\degree$ is $\lesssim$$4\%$, assuming uniform $\ostar$ in $[0,180]\degree$ and uniform $\istar$ on the sphere.
Overall, the probability of stellar flybys, that can satisfy all four constraints derived from the observed sednoid population, is smaller than $5\%$.
The observational constraints we summarize here, if confirmed by future sednoid discoveries, serve as a powerful way to test different sednoid formation hypotheses.

If we remove the constraint of the primordial orbital alignment, which arguably needs further observations to confirm \citep[e.g.,][]{2025NatAsChen}, we have 71 out of 768 stellar flybys left, among which 31 are for extremely close ($\qstar = 300,500$ au) stellar encounters with field stars ($\vstar = 10,30$ km/s), and 40 are for close stellar encounters within the Sun's birth cluster ($\vstar=1,3$ km/s). The probability of the former 31 flybys is $\lesssim 1\%$ as argued before. After examining the latter 40 cases, we find that these flybys either have $\istar\sim0\degree$ or $\ostar\sim0\degree$ due to the low-inclination constraint, and the probability of such flyby geometry is $\lesssim$$20\%$.

\section{Discussion}

\subsection{Alternatives to Stellar Flyby}
Beyond the stellar flyby scenario, multiple competing hypotheses exist for forming sednoids---most notably the ``rogue planet" and ``Planet Nine" models---which we now evaluate through the lens of our newly identified constraints.

The ``rogue planet" hypothesis posits that a transient planetary-mass perturber in the early Solar System could have populated the primordial scattering disk efficiently \citep{Gladman.2006, Huang.2022ml}.
\citet{Huang.2024s} demonstrates that a $2M_\oplus$ rogue planet, surviving for $\sim$100 Myr before being ejected by Neptune, can generate sednoid populations spanning from $\sim$200 to a few thousand au in semimajor axis.
Crucially, this mechanism produces sednoids and eTNOs with relatively confined inclination distributions, compatible with our $\lf>50\%$ constraint.
During its $\sim$100 Myr temporary presence, the rogue planet creates strong \textit{primordial} clustering in the longitudes of perihelion with circular standard deviation $\sigma_\varpi \approx 25^\circ$ (corresponding to $\kappa\approx 5$) for $q > 50$~au populations, which, after the rogue planet's ejection, was dispersed by differential precessions induced by the four giants.

The ``Planet Nine" or other existing planet hypothesis posits that an undiscovered Earth-sized planet lurks at several hundred au and continuously sculpts the orbits of distant TNOs (see review by \citealt{PN_Batygin2019}). Numerous simulations have investigated the dynamical evolution of TNOs under the perturbation of an existing distant planet and found that eTNOs, including sednoids, can be efficiently generated \citep{PNsim_Lykawka2008,PNsim_Batygin2016,PNsim_Lykawka2023}.
However, the eTNOs generated by an existing planet exhibit relatively high inclinations (see \citealt{Batygin.2016}), even on retrograde orbits, potentially inconsistent with the low-$i$ constraint derived in this work (see details in Section \ref{subsect:low-i}).
Besides, ``Planet Nine" hypothesis predicts \textit{current} orbital clustering, rather than \textit{primordial} orbital alignment, among distant TNOs, which is testable with well-characterized surveys.

Although the present sample of sednoids is still insufficient to distinguish between different formation channels at an irrefutable statistical level, the Vera Rubin Observatory (or LSST, \citealt{Schwamb.2023}) is expected to discover several more and offer a definitive opportunity to examine the constraints we propose here.

\subsection{Retrograde Sednoids?}
In a recent exploration of the stellar flyby scenario, \citet{Pfalzner.2024} proposed that to produce the ``observed TNO properties'', the flyby was likely a $0.8$~$M_\odot$ star passing at an extremely close distance with $q_\star = 110$~au on a nearly vertical orbit $i_\star = 70^\circ$. We find this scenario implausible for two critical reasons: (1) the stability of planetary orbits and the unheated nature of the cold classical belt both require that a passing star must maintain $q_\star \gtrsim 240$~au \citep{adams2010birth,Batygin.2020jr}; and (2) the low-inclination distribution of eTNOs, including the $\lf < 50\%$ constraint we derived in this study, would be violated.

Besides, \citet{Pfalzner.2024} based their parameter selection on 2019 EE$_6$, a purported ``retrograde sednoid''\footnote{This should not be confused with retrograde Centaurs and TNOs coupled with Neptune ($q \lesssim$ 30~au), such as 2008 KV$_{42}$ \citep{Gladman.2009} and 471325 Taowu (2011 KT$_{19}$, \citealt{Chen.2016}). They are a real population and could originate from the Oort Cloud (see \citealt{Brasser.2012aml} and \citealt{Ito.2024}).}. With a data arc span of 63 days, the orbit of 2019 EE$_6$ has enormous uncertainty, with a $1\sigma$ error in inclination of $\sim$$120^\circ$ according to JPL Small Body Database\footnote{\url{https://ssd.jpl.nasa.gov/tools/sbdb_lookup.html}}.
Apart from 2019 EE$_6$, other two objects (2022 FM$_{12}$ and 2022 FN$_{12}$) were also claimed to be ``regrograde sednoids", based on short observing baselines and thus large orbital uncertainties \citep{Migaszewski.2023}. However, 2022 FM$_{12}$ and 2022 FN$_{12}$ have recently been revised to be ordinary cold classical TNOs on low-$e$ and low-$i$ orbits (according to JPL Small Body Database), which necessitates more observations to further constrain the orbit of 2019 EE$_6$.
Fortunately, the upcoming LSST survey with a 10-year baseline is promising to put more stringent constraints on the existence of retrograde sednoids.

\section{Conclusion}
Our dynamical simulations demonstrate that single stellar flyby scenarios are unlikely ($\lesssim5\%$) to reproduce the observed sednoid population in terms of four key aspects: the efficiency of sednoid implantation with a minimum rate of $\eta > 5\%$, the coverage of semimajor axes encompassing all observed sednoids (especially 2012 VP$_{113}$), the low-inclination dominance as indicated by both sednoid and eTNO samples ($f_{i<30^\circ} > 50\%$), and the strong primordial apsidal clustering as inferred from the three sednoids (Sedna, 2012 VP$_{113}$, and Leleakuhonua). These constraints, particularly the low-inclination preference and the primordial orbital alignment, establish a new benchmark for evaluating outer Solar System formation models.


\section{Acknowledgments}
QH thanks Helong Huang for useful discussion regarding the sednoid implantation efficiency. We would like to thank the anonymous referee for comments on the previous manuscript. We thank Simon Portegies Zwart for useful discussions. Work by QH, YH, and WZ was supported the National Natural Science Foundation of China (grant Nos. 12173021 and 12133005). YH acknowledges support by JSPS KAKENHI No. 25K17460.

%

\vspace{5mm}


\software{\texttt{rebound} \citep{rebound},
          \texttt{scipy} \citep{2020SciPy-NMeth},
          \texttt{numpy} \citep{numpy},
          \texttt{matplotlib} \citep{matplotlib}
          }



\appendix
\section{Discovery Surveys of eTNOs}\label{appen:1}
\begin{table}[htbp]\label{tab:survey}
    \caption{Information of the 19 eTNOs used in this study.}
    \begin{center}
    \begin{tabular}{c|llll}
    \hline\hline
    & Name & Discovery Survey & Discovery Paper & $i\ (\degree)$ \\
    \hline
    1 & 2000 CR$_{105}$ & Deep Ecliptic Survey & \citet{2000CR105} & 22.74 \\
    2 & 2013 RA$_{109}$ & Dark Energy Survey & \citet{DESfirstyear} & 12.39 \\
    3 & 2014 WB$_{556}$ & Dark Energy Survey & \citet{DESfirstyear} & 24.15 \\
    4 & 2016 SD$_{106}$ & Dark Energy Survey & \citet{Bernardinelli.2022} & 4.8 \\
    5 & 2004 VN$_{112}$ & ESSENCE Supernova Survey & \citet{2004VN112} & 25.49 \\
    6 & 2010 GB$_{174}$ & Next Generation Virgo Cluster Survey & \citet{2010GB174} & 21.56 \\
    7 & 2013 UT$_{15}$ & Outer Solar System Origins Survey & \citet{OSSOS2018DR} & 10.63 \\
    8 & 2013 SY$_{99}$ & Outer Solar System Origins Survey & \citet{Bannister.2017} & 4.22 \\
    9 & 2015 KG$_{163}$ & Outer Solar System Origins Survey & \citet{OSSOS2018DR} & 14.02 \\
    10 & 2015 RX$_{245}$ & Outer Solar System Origins Survey & \citet{OSSOS2018DR} & 12.12 \\
    11 & 2013 FT$_{28}$ & --- & \citet{2013FT28} & 17.42 \\
    12 & 2014 SR$_{349}$ & --- & \citet{2013FT28} & 17.95 \\
    13 & 2018 VM$_{35}$ & --- & \citet{Sheppard.2019} & 8.47 \\
    14 & 2019 EU$_{5}$ & --- & \citet{Sheppard.2019} & 18.17 \\
    15 & 2021 DK$_{18}$ & --- & \citet{Sheppard.2019} & 15.42 \\
    16 & 2021 RR$_{205}$ & --- & \citet{Sheppard.2019} & 7.65 \\
    \hline
    17 & Sedna & --- & \citet{Brown.2004} & 11.90 \\
    18 & 2012 VP$_{113}$ & --- & \citet{Trujillo.2014} & 24.10 \\
    19 & Leleakuhonua & --- & \citet{Sheppard.2019} & 11.70 \\
    \hline\hline
    \end{tabular}
    \end{center}
    \tablenotetext{}{Note: The orbital inclinations were retrieved from JPL Small-Body Database on January 20th, 2025.}
\end{table}

The fact that all $q > 38$~au eTNOs discovered to date have $i < 30\degree$ has not received sufficient attention as a constraint on outer Solar System formation models. This is partly due to the misconception that most eTNO discoveries come from surveys limited to the ecliptic plane, leading to the attribution of the observed $i < 30\degree$ distribution solely to inclination bias.


In Table~\ref{tab:survey}, we summarize the discovery surveys and orbital inclinations of the 19 detached eTNOs used in Section~\ref{subsect:low-i}. Only 5 of the 19 were discovered by ecliptic surveys, including OSSOS (see Figure~2 in \citealt{OSSOS2018DR} for sky coverage) and the Deep Ecliptic Survey (see Figure~1 in \citealt{2005AJDES}). By contrast, 14 of the 19 eTNOs in our sample were found by non-ecliptic surveys: the ESSENCE supernova survey ($-22\degree$ to $-5\degree$ in ecliptic latitude; Table~1 in \citealt{2007ApJESSENCE}), the Next Generation Virgo Cluster Survey (NGVS; centered at $+13\degree$; Figure~1 in \citealt{2012ApJSNGVS}), the Sheppard \& Trujillo sednoid/eTNO survey (covering $\pm30\degree$; Figure~2 in \citealt{Sheppard.2019} for all their pointings), the Dark Energy Survey (DES; mostly $>30\degree$; Figure~1 in \citealt{Bernardinelli.2022}), and the all-sky survey by Mike Brown that discovered Sedna \citep{Brown.2004}.

Although it is nearly impossible to fully remove observational biases in eTNO orbital inclinations due to the heterogeneity of discovery surveys, the fact that three quarters of the known eTNOs were found by non-ecliptic surveys suggests that the prevalence of low-$i$ eTNOs likely reflects the intrinsic population and therefore requires explanation.

\section{The Statistics of All Stellar Flyby Simulations}

\begin{table*}[]
    \caption{The statistics of all stellar flyby simulations.}
    \label{tab:statistics}
    \begin{center}
    \begin{tabular}{cccccccccccccc}
    \hline\hline
    $\mstar\,(\msun)$ & $\vstar$ (km/s) & $\qstar$ (au) & $\istar\,(\degree)$ & $\ostar\,(\degree)$ & Con1 & Con2 & Con3 & Con4 & $\eta$ & $a_{\text{sim}}^{2.5\%}$ (au) & $a_{\text{sim}}^{97.5\%}$ (au) & $\lf$ & $\kappa_\text{sim}$ \\
    \hline
    0.3 & 1 & 300 & 0 & 0 & 1 & 1 & 0 & 0 & 0.20 & 217.7 & 1405.7 & 0.47 & 0.63 \\
    0.3 & 1 & 300 & 0 & 45 & 1 & 1 & 0 & 0 & 0.20 & 222.5 & 1411.7 & 0.47 & 0.63 \\
    0.3 & 1 & 300 & 0 & 90 & 1 & 1 & 0 & 0 & 0.20 & 216.5 & 1395.9 & 0.47 & 0.69 \\
    0.3 & 1 & 300 & 0 & 135 & 1 & 1 & 0 & 0 & 0.20 & 216.9 & 1388.9 & 0.47 & 0.66 \\
    0.3 & 1 & 300 & 36 & 0 & 1 & 1 & 0 & 0 & 0.22 & 217.4 & 1383.2 & 0.39 & 1.29 \\
    ... &&&&&&&&&&&&& \\
    1.0 & 1 & 300 & 0 & 0 & 1 & 1 & 1 & 0 & 0.21 & 207.5 & 1304.3 & 0.63 & 0.92 \\
    1.0 & 1 & 300 & 0 & 45 & 1 & 1 & 1 & 0 & 0.21 & 210.4 & 1328.0 & 0.61 & 1.09 \\
    1.0 & 1 & 300 & 0 & 90 & 1 & 1 & 1 & 0 & 0.22 & 209.3 & 1324.9 & 0.63 & 1.04 \\
    1.0 & 1 & 300 & 0 & 135 & 1 & 1 & 1 & 0 & 0.21 & 209.2 & 1308.1 & 0.64 & 1.06 \\
    1.0 & 1 & 300 & 36 & 0 & 1 & 1 & 0 & 0 & 0.23 & 212.7 & 1330.6 & 0.43 & 1.26 \\
    ... &&&&&&&&&&&&& \\
    \hline
    \end{tabular}
    \end{center}
    \tablenotetext{}{Note: The whole table is available as a machine-readable one in the online manuscript.}
\end{table*}

We present the statistics for all 768 stellar flyby simulations in Table~\ref{tab:statistics}, enabling readers to readily identify which flybys remain viable under specific constraints. Here, ``Con1'' through ``Con4'' correspond to the ``Implantation Efficiency'', ``Semimajor Axis Range'', ``Low-inclination Fraction'', and ``Primordial Clustering Range'' constraints, respectively. An entry of ``1'' in these columns indicates that the stellar flyby satisfies the corresponding constraint, whereas an entry of ``0'' indicates that it does not. All other symbols are defined consistently with those in the main text.



\bibliography{references}{}
\bibliographystyle{aasjournal}


\end{CJK*}
\end{document}